\documentclass[12pt,preprint]{aastex}
\begin{document}
\pdfoutput=1
\shorttitle{SGRBs in dense stellar systems}

\shortauthors{Parsons, Ramirez-Ruiz \& Lee}

\title{Short Gamma Ray Bursts and their Afterglow Signatures in Dense
  Stellar Systems}

\author{Rion K. Parsons\altaffilmark{1}, Enrico
  Ramirez-Ruiz\altaffilmark{1} and William H. Lee\altaffilmark{2}}
\altaffiltext{1}{Department of Astronomy and Astrophysics, University
  of California, Santa Cruz, CA 95064}\altaffiltext{2}{Instituto de
  Astronom\'{\i}a, Universidad Nacional Aut\'{o}noma de M\'{e}xico,
  Apdo. Postal 70-264, Cd. Universitaria, M\'{e}xico D.F. 04510}

\begin{abstract} %
The hypothesis that short GRBs arise from the coalescence of binary
compact stars has recently gained support. With this comes the
expectation that the afterglow should bear the characteristic
signature of a tenuous intergalactic medium (IGM). However, fits to
the observational data suggest that some detected afterglows arise in
relatively dense gaseous environments rather than in the low density
IGM. Here we show that considering the effect of red giant winds in
the core of a star cluster may resolve this paradox if short GRB
progenitors are contained in such an environment and close encounters
rather than pure gravitational wave emission brings the compact
objects together. Clear confirmation is provided here of the important
notion that the morphology and visibility of short gamma-ray burst
remnants are determined largely by the state of the gas in the
cluster's core.
\end{abstract}


\section{Introduction}
Although they were discovered roughly forty years ago
\citep{klebesadel73}, Short $\gamma$-Ray Bursts (SGRBs) are still a
mystery. All that can confidently be said is that they involve compact
objects and highly relativistic dynamics.  The most widely favored and
conventional possibility is that they are produced by the coalescence
of compact object binaries involving neutron stars and/or black holes,
although other alternatives should also be considered \citep[see
  e.g.][for reviews]{nakar07,lrr07}. Such systems, which are known to
exist \citep{ht75,burgay03}, are driven to coalesce by energy and
angular momentum losses to gravitational radiation.  In this scenario,
the compact binary would take hundreds of millions of years to spiral
together, and could by then (especially if given a kick velocity on
formation) have moved many kilo-parsecs from the site of its birth
\citep{fryer99,bloom99}. However, fits to the observational data
suggest that some detected afterglows
\citep{berger07,nakar07,panaitescu06,nfp08} arise in relatively dense
gaseous environments rather than in the low density Intergalactic
Medium (IGM). We must therefore remain aware of other
possibilities. It may be wrong, for instance, to suppose that the
binary is formed in isolation, since it could instead be the result of
an encounter in a dense stellar system.

Most stars in the Universe never interact strongly with others, at
least during their adult life, after they have left the interstellar
gas cloud in which they were born. However, there are various {\it
  dense stellar systems} such as star clusters and galactic centers,
where stars are sufficiently close to their neighbors to make
encounters significantly more likely. To get a sense of how crowded
such a region is, consider the contents of the inner core of a
Globular Cluster (GC).  While in the Solar neighborhood the typical
distance between individual stars is more than 1~pc, a sphere with a
radius of 0.1~pc around the core of a GC contains over a hundred
thousand, if not millions, of solar masses in stars following tight
orbits about the nucleus. The stellar density is therefore more than a
million times higher than that in the neighborhood of our Sun. In such
environments, it is unavoidable that many single and binary stars will
undergo close encounters and even physical collisions within their
lifetimes.  Indeed, it is thought that this type of interaction can
lead to the formation of blue stragglers \citep{glebbeek08}, and
possibly intermediate mass black holes \citep{portegies04}.  It is in
these environments, called dense stellar systems, that a compact
object binary can be formed by two- and three-body encounters
\citep{grindlay06}.  The subsequent merger, or possibly direct
collisions of compact objects\citep{lrr07}, can produce a SGRB, and
the resulting afterglow could then at least in part be due to the
interaction of the relativistic ejecta with the stellar winds of the
cluster members.  Due to the large stellar density in the star cluster
core, the interaction of the external shock can take place with a
denser external medium than that of the IGM.  Much of the effort
herein will be dedicated to determining the state of the circumburst
material in the cores of GCs, and describe how this external matter
can affect the observable burst and afterglow characteristics.

\section{SGRB\lowercase{s} from Dense Stellar Systems}

First, we assume that different types of stellar objects are
distributed homogeneously in a spherical core of radius $r_{\rm
  c}=0.1r_{\rm c,-1}$ pc, within which the number density, $n_{\rm
  c}$, and stellar velocity dispersion, $\sigma_{\rm c}$, are
constant.  The gas density in the cores of GCs depends on the various
types of stellar members that populate their interiors.  In a typical
cluster, winds from red giant stars tend to have the greatest effect
on the core's gas density despite being significantly outnumbered by
other stellar types.  These winds are typically slow-moving and dense,
with velocities on the order of $v_{\rm w}= 10 v_{\rm w,1}$ km
s$^{-1}$ and mass loss rates between $10^{-7}$ and $10^{-6}$
M$_{\odot}$ yr$^{-1}$.  In a steady, spherically symmetric solution,
the electron density is
\begin{equation}
n_w(r)\approx 3\times 10^{3} r_{16}^{-2} v_{\rm w,1}^{-1}\dot{M}_{\rm
  w, -7} \mu_e^{-1}\,{\rm cm}^{-3},
\end{equation}
where $\mu_e\sim 2$ in a Helium gas, $r=10^{16}r_{16}$ cm and
$\dot{M}_{\rm w}=10^{-7}\dot{M}_{\rm w,-7}$ M$_{\odot}$ yr$^{-1}$.
The inner core of a GC contains $N_\ast=10^2N_{\ast,2}$ red giants,
and so their mean separation is
\begin{equation}
r_\perp= 6.4 \times 10^{16} N_{\ast,2}^{-1/3} r_{\rm c,-1}\;{\rm cm}.
\end{equation}		
We can thus estimate a minimum average density $n_{\perp}$ in the
cores of GCs in the absence of gas retention as:
		\begin{equation}
			n_\perp \sim n_{\rm w} (r_\perp) =80
                        N_{\ast,2}^{2/3}r_{\rm c,-1}^{-2} v_{\rm
                          w,1}^{-1} \dot{M}_{\rm w, -7}
                        \mu_e^{-1}\,{\rm cm}^{-3}.
		\end{equation}		
In this case, the SGRB would expand into a medium that is
significantly denser than the IGM.

For this discussion we will assume that the blast wave is adiabatic,
i.e. its energy is constant with time, and effectively spherically
distributed.  This means that the energy ($E=10^{50}E_{50}$ erg) is
the {\em isotropic equivalent} energy as, for example, derived from
the $\gamma$-ray output.  Deceleration due to the combined stellar
winds starts in earnest when about half the initial energy has been
transferred to the shocked matter, i.e. when it has swept up
$\Gamma^{-1}$ times its own rest mass.  The typical mass where this
happens is $M_{\rm dec}=E/(\Gamma^{2}c^2) \approx 5 \times 10^{-8}
E_{51} \Gamma_{2}^{-2} M_\odot$. In the GC core, the mass within
radius $r$ is ${4 \pi \over 3}\rho_{\perp} r^3$, which gives the blast
wave deceleration radius
\begin{equation}			
r_{\rm d}=10^{16} E_{50}^{1/3}\Gamma_2^{-2/3} \Big({n_\perp
  \over 1\;\rm{cm}^{-3}}\Big)^{-1/3}\; {\rm cm}.
\label{eqn:rdec}
\end{equation}		
A blast wave in a GC core thus decelerates at a much smaller radius
than it would in the IGM. 

Under the assumption that energy conversion takes place primarily
within the forward shock, the energy equation reads $E_{\rm
  bw}(t)=\psi \Gamma^{2}M_{\rm bw}c^{2}$, where $E_{\rm bw}$ is the
blast wave's isotropic equivalent energy, $\psi$ is a constant of
order unity \citep{bm76}, and $M_{\rm bw}(r_{\rm bw}) \equiv 4\pi \int
_{0}^{r_{\rm bw}}n_{\rm w}(r)r^{2}dr$ is the cumulative swept-up
mass. The observed time interval during which most of the photons
emitted (at a radius $r_{\rm bw}$) are received is of the order of
$r_{\rm bw}/(4\Gamma^2c)$ \citep{spn98}.

As usual, we assume that the dominant radiation process is synchrotron
emission. If the energy in the magnetic field and electrons are taken
to be a fraction $\epsilon_{\rm B}$ and $\epsilon_{\rm e}$,
respectively, of the thermal energy density, then approximating the
electron energy distribution as a power law with an index $p$, the
afterglow flux during the adiabatic expansion of the blast wave is
given by
\begin{equation}
F_{\nu }\propto \nu^{(1-p)/2} n^{(1+p)/4}E_{\rm bw}^{p}M_{\rm
  bw}^{1-p}{\epsilon}_{\rm B}^{(1+p)/4}{\epsilon}_{\rm e}^{(p-1)}
\label{eqn:fu1}
\end{equation}
for $\nu_{\rm sy}<\nu <\nu_{\rm c}$, and
\begin{equation}
F_{\nu }\propto \nu^{-p/2}n^{(p-2)/4}E_{\rm bw}^{p-1}M_{\rm
  bw}^{2-p}t^{-1}{\epsilon}_{\rm B}^{(p-2)/4}{\epsilon}_{\rm
  e}^{(p-1)}
\label{eqn:fu2}
\end{equation}
for $\nu_{c}<\nu$. The above relations are valid for varying energy or
density, as can be seen by considering the implicit time dependence
through $E_{\rm bw}(t)$, $n(t)$ and $M_{\rm bw}(t)$.

When the dominant variations are in the circumburst matter rather than
energy, one may write $t=(c/4E)(M_{\rm bw}r_{\rm bw} + \int_0^{r_{\rm
    bw}} r^2n(r)dr)$. Under this assumption, equations (\ref{eqn:fu1})
and (\ref{eqn:fu2}) reduce to $F_\nu \propto M_{\rm bw}^{1-p}
n^{(p+1)/4}$ and $F_\nu \propto M_{\rm bw}^{2-p}t^{-1}\;n^{(p-2)/4}$,
respectively. As the blast wave expands into the cluster winds, strong
temporal variations compared to the canonical power-law decay can thus
be produced as a result of changes on the properties of the star
cluster, especially the mass-loss rates of the stars and their number
density.  The characteristic mass scale where this takes place is $
M_{\rm bw} \gtrsim M_{\rm dec}$. This phase ends when so much mass
shares the energy that the $\beta \Gamma \leq 1$, setting a
non-relativistic mass scale $M_{\rm NR}=E/c^2 \approx 5 \times 10^{-5}
E_{50} M_\odot$.  Beyond this point, the event slowly changes into a
classical Sedov-Taylor supernova remnant evolution.

In the absence of characteristic scales in stellar ejecta and in the
ambient medium, self-similar, spherically symmetric solutions exist,
and they are widely used to interpret observational data on
afterglows.  By contrast, the interaction of a GRB with a non-uniform
medium is poorly understood. The presence of a density gradient will
only affect the dynamics of the GRB when the remnant size is
comparable to, or exceeds, the scale length of the gradient. Before
this time, the density can be treated as approximately uniform.  Thus,
it is only when $r_{\rm d} \gtrsim r_\perp$ that the details of the
transition are important for the dynamics. This requires
\begin{equation}
{E_{51} v_{\rm w,1}  r_{\rm c,-1} \mu_e \over \Gamma_2^{2}N_{\ast,2}^{1/3}  \dot{M}_{\rm w, -7}} \gtrsim 10^{3}.
\label{eqn:rperprd}		
\end{equation} 
A close examination of equation (\ref{eqn:rperprd}) shows that if the
winds of red giants are especially weak (i.e. $\dot{M}_{-7} < 1$) or
the burst's energy content is large ($E_{51} > 1$), $r_{\perp}$ falls
within the range of relativistic expansion. Otherwise, this radius is
sufficiently large that the interaction with the free wind of the
nearest red giant is expected over the typical period of observation
of afterglows.

Depending on the wind properties of the stellar members as well as
their number density, however, the density structure in this region
could be quite complicated as the winds of stars interact at a typical
distance $\sim r_\perp/2$.  Each free expanding wind encounters an
inward facing shock, where the typical preshock densities are $\sim 4
n_{\perp} \sim 320 \dot{M}_{\rm w,-7}N_{\ast,2}^{2/3}r_{\rm
  c,-1}^{-2}v_{\rm w,1}^{-1}\;{\rm cm}^{-3}$.  Kinetic energy is
deposited in the shocked wind region in the form of heat, with
temperature $T_{\rm shock} =(3m_{\rm p}\mu_{\rm e}/16k) (\Delta v_{\rm
  w})^2=4\times 10^3(\Delta v_{\rm w}/10\;{\rm km\;s^{-1}})^2 \;{\rm
  K}$, where $\Delta v_{\rm w}$ is the speed of the material relative
to the approaching shock.

The shock interactions between nearby stars will be radiative if the
cooling distance $r_{\rm cool}$ satisfies the condition $\kappa=r_{\rm
  cool}/r_\perp<1$. Using the cooling distances obtained by
\citet{hartigan87} for shocks in the low-density regime derived from
self-consistent pre-ionization, plane-parallel, steady models, we
obtain $\kappa\ll 1$. A cluster wind in this high cooling regime will
have dense, cool structures between nearby stars and the SGRB blast
wave could interact with them while still relativistic provided that
$r_{\rm d} < r_\perp$. Large-scale density inhomogeneities in the
circumburst medium are, however, likely to result in distortions of
the ejecta that might be more readily observed as the blast wave
evolves into the non-relativistic phase.

\section{Cluster Winds Interaction Models for SGRB\lowercase{s}}

\subsection{Initial Model}
\begin{figure}
\plotone{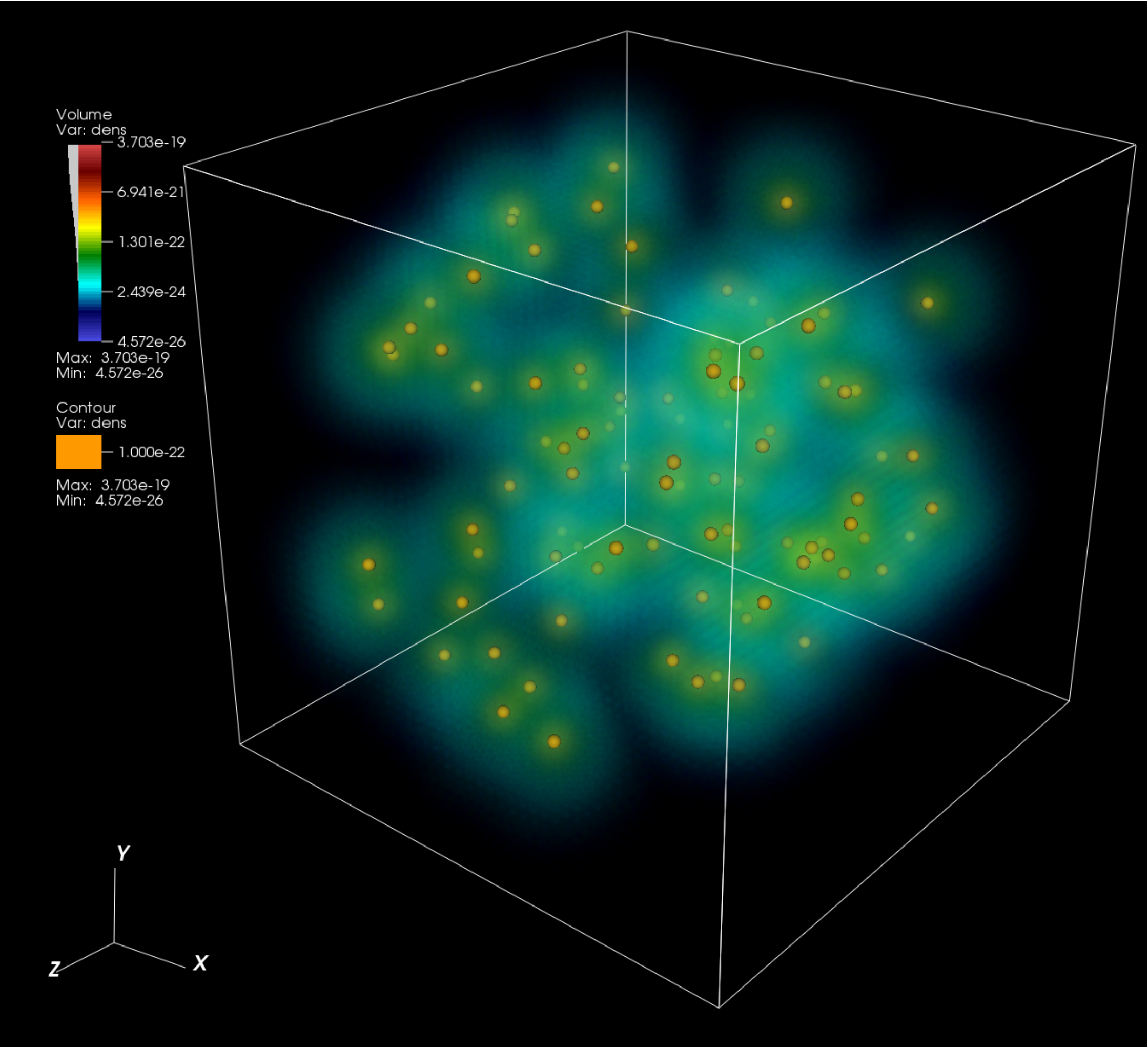}
\caption{Simulated red giant stars within the GC core. Each star is
  surrounded by a $1/r^{2}$ density profile, and is in ram pressure
  equilibrium with neighboring stars. The yellow surface is a contour
  of constant density (in $M_\odot c^{-3} s^{-3}$ units) around the
  star, and is not representative of the actual (unresolved) stellar
  radius.  The size of the computational domain is (0.2 pc)$^{3}$. }
\label{fig1}
\end{figure}

To build the initial model, $N_\ast$ stars were randomly distributed
in a three-dimensional volume such that the mean separation between
stars is r$_{\perp} \sim N_{\ast}^{1/3} r_{\rm c}$.  The computational
box is $8.0 \times 10^{17}$ cm in each dimension with the stars
located within the central $r_{\rm c} \sim 6.0 \times 10^{17}$ cm.
Based on the data from M15 \citep{dull97,vandenbosch2006}, the
number density of red giants within r$_{\rm c}$ is somewhere between
10 and 100, and for this particular simulation we set $N_\ast=10^2$,
such that $r_{\perp} \sim 6 \times 10^{16}\;\rm{cm}$.  Each star was
given the same effective mass loss rate $\dot{M}_{\rm RG} = 10^{-7}$
M$_{\odot}$ yr$^{-1}$ and stellar wind speed v$_{\rm RG} = 10$ km
s$^{-1}$.  Based on $\dot{M}_{\rm RG}$ and v$_{\rm RG}$, a spherical
$1/r^2$ wind profile was implemented for each cluster member (Figure
\ref{fig1}).  The individual density profiles were then extrapolated
so that they are in ram pressure equilibrium with the closest
neighbor.

\subsection{Dynamics of SGRB Remnants}
In the case depicted in Figure~\ref{fig1} the stellar winds of the
individual members are dense enough to slow down the ejecta (for
reasonable SGRB properties) to non-relativistic speeds before reaching
$r_\perp$, so that we expect the blast wave evolution as we see it to
take place in the free-expansion phase. This encourages us to present
a detailed account of resulting dynamics of the remnants within a
dense stellar system following the onset of the non-relativistic
phase.  Common to all calculations is the initiation of the SGRB
explosion as two identical blobs expanding in opposite directions into
the circumburst medium. Calculations were done in three dimensions
using the PPM adaptive mesh refinement code FLASH (ver 2.5). The blobs
and the circumburst medium are modeled as a cold ideal gas with
$\gamma = 5/3$.  The numerical domain is a unprolonged cylinder in
which the ejecta moves along the {\it y}--axis. In the inner region of
each of the pancakes, the ejecta mass, $M_{\rm j}$ is distributed
uniformly and for all runs we have used $\Gamma_{\rm j}=2$, $\Delta
r_{\rm j}/r_{\rm j}\sim 0.6$, $\theta_{\rm j} \sim 0.5$, and $r_{\rm
  j}\sim r_{\rm d} (\Gamma_{\rm j})$.

Without a detailed understanding of the exact shape and energy
distribution of the ejecta, we have only an approximate description of
how to construct the initial conditions. However, as clearly
illustrated by \citet{ap01} and \citet{rrm08}, the late time evolution
of the ejecta is rather insensitive to uncertainties in the initial
conditions. We have considered various initial densities, angular
widths, and shapes of the collimated ejecta and found that these are
indeed unimportant in determining the late morphology of the
remnant. This stems from the fact that at late times the mass of the
remnant is dominated by the circumburst gas, which washes out any
variations in the initial conditions of the ejecta.

\begin{figure}
\plotone{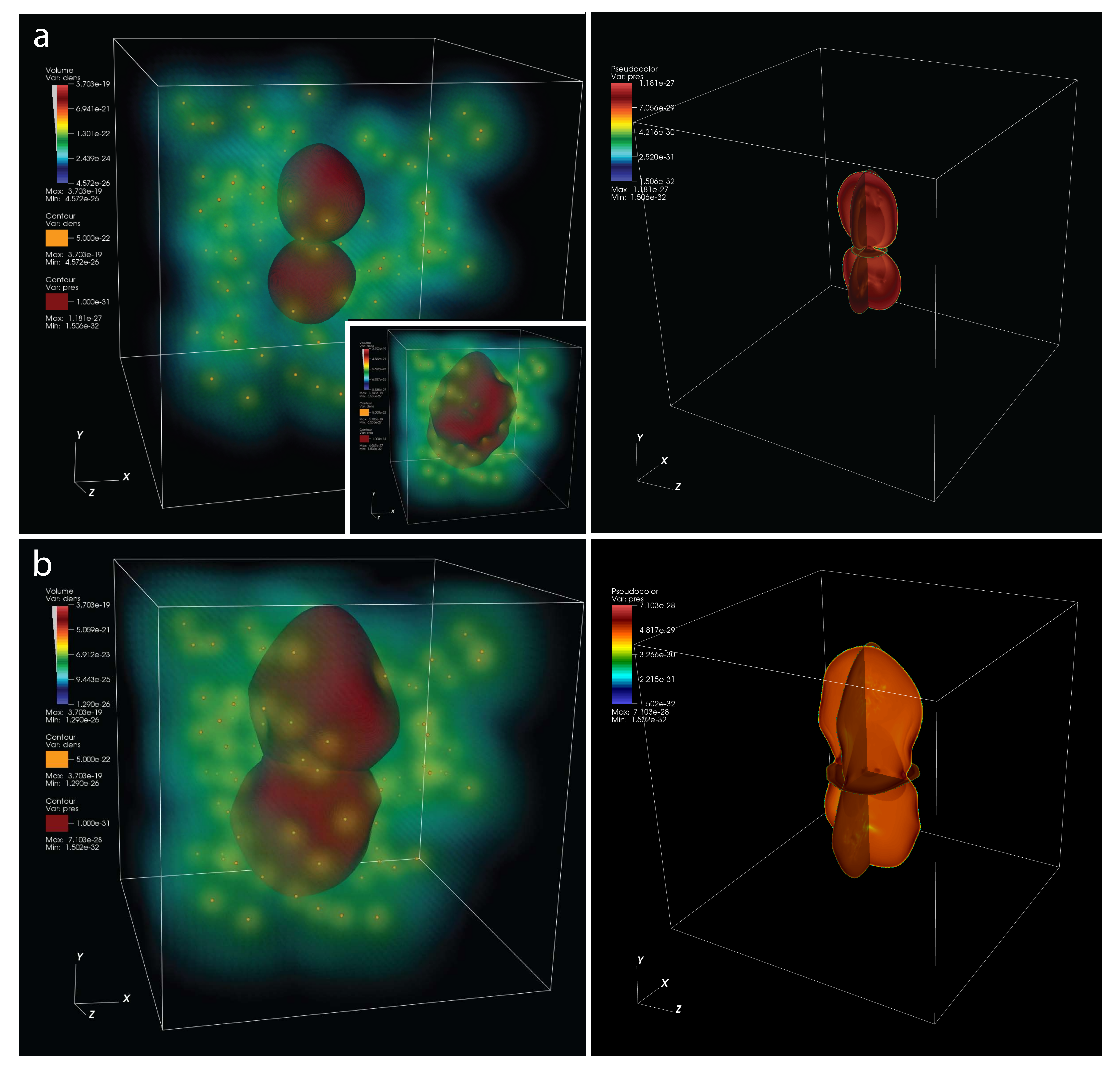}
\caption{Morphology of the SGRB remnant $4.8\times 10^7$ (a, top row)
  and $1.9 \times 10^8$ (b, bottom row) seconds after explosion. In
  the left panels, the pressure isosurface (in units of $M_\odot
  c^{-1} s^{-3}$) is located at the outer boundary of the blast wave,
  and yellow/green/blue shading shows the density profile around the
  stars (in $M_\odot c^{-3} s^{-3}$ units). In the early stages, the
  blast wave remains largely undisturbed. The size of the
  computational domain is (0.2 pc)$^{3}$, and calculations were
  carried out in cartesian coordinates in three dimensions with six
  levels of refinement. The pressure slices in the right column show
  the region interior to the blast wave for each case. {\it Inset
    panel:} Behavior of a spherical explosion with a comparable
  isotropic equivalent energy $E_{\Omega}=5 \times 10^{48}$ erg.  The
  initial high pressure region is in this case is a sphere in which
  the ejecta mass, and the thermal energy are initially uniformly
  distributed.}
\label{fig2}
\end{figure}

Detailed hydrodynamic simulations of the evolution of a GRB remnant in
the GC core medium are presented in Figure \ref{fig2}, where the
pressure contours of the expanding ejecta are plotted.  As the gas
collides with the external medium, a blast wave forms that propagates
in the radial direction of motion and, over time, wraps around the
various density discontinuities produced by the red giant stars.  As
it can be seen in panel (a) in Figure \ref{fig2}, the evolution of the
blast wave suffers only minor distortions as the remnant sweeps a mass
$\sim E/c^2$. By the time the blast wave reaches the edge of the
cluster's core, it becomes fairly distorted, though it still holds a
jet-like appearance.  Pronounced dimples are clearly visible in the
lower portion of the blast wave showing where stars have been
overtaken. Also shown in Figure~\ref{fig2} for comparison is the
evolution of a spherical blast wave of comparable isotropic equivalent
energy.

\begin{figure}
\epsscale{0.5}
\plotone{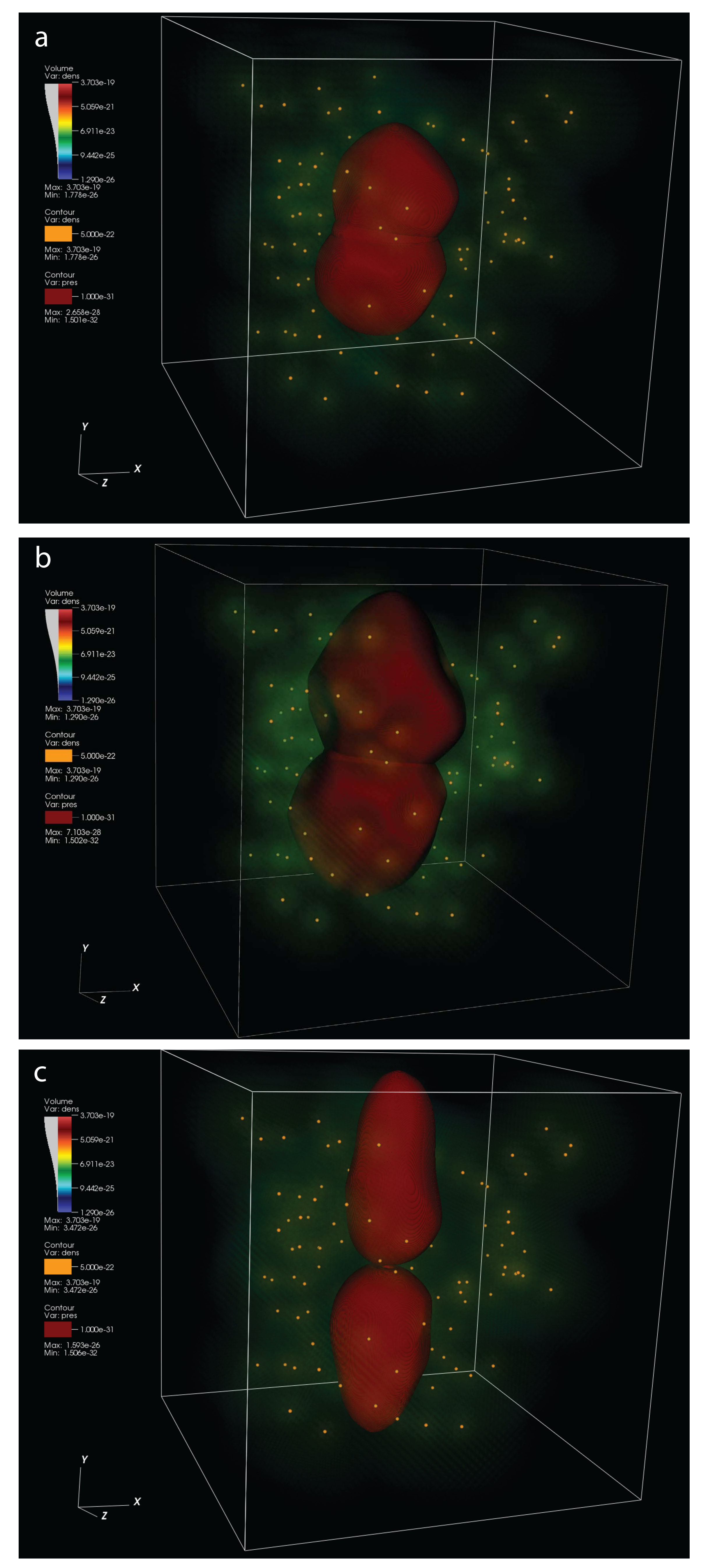}
\caption{The evolution of a SGRB remnant with varying energy at $t=2.3
  \times 10^8$, $10^8$ and $4.4 \times 10^7$ seconds (panels a,b and
  c, respectively). Shown are 3D pressure isosurfaces in normalized
  units ($M_\odot c^{-1} s^{-3}$) for calculations with $E_{\Omega}=5
  \times 10^{47}$, $5 \times 10^{48}$ and $5 \times 10^{49}$ erg
  (panels a,b and c, respectively).  The blast wave has reached the
  edge of the core radius, at which point it has become extremely
  distorted.  Pronounced dimples are clearly visible in the lower
  portion of the blast wave showing where stars have been overtaken. }
\label{fig3}
\end{figure}

The calculations above demonstrate how the dynamical evolution of SGRB
remnants in dense stellar systems depends sensitively on the stellar
wind properties of the cluster members as well as their density
distribution. This confirms the notion that that the morphology and
visibility of short gamma-ray burst remnants are determined largely by
the state of the gas in the cluster's core. The resulting evolution
depends also fairly strongly on the properties of the GRB ejecta,
especially its energy content, since it sets the non-relativistic mass
scale for a particular stellar cluster. Figure \ref{fig3} illustrates
the dependence on energy of the remnant's morphology, where structures
similar to those described in Figure \ref{fig2} are clearly seen. A
bow shock forms as each blob collides with the surrounding medium,
which eventually wraps around the ejecta before the two expanding
shells collide to form a single structure. However, in the low energy
content case (panel a) the GRB remnant decelerates much more rapidly
and although initially it may be highly nonspherical, the aspect ratio
approaches unity as the two blobs expand and merge before reaching the
edge of the GC core at about $2.3 \times 10^8$ seconds.  Beyond this
point, the evolution will follow the  classical Sedov-Taylor
supernova remnant solution.  With higher energy content, on the other
hand, deceleration to non-relativistic velocities occurs at much
larger radii (panel c), and the blast wave is still moving through the
core region before the two expanding shells collide to form a single
structure.

\section{Discussion}
It is evident from the above discussion that the environment in the
cores of GCs is a very rich one in terms of observable
consequences. Even in the simplest case of red giant stars with
identical wind properties, complex behavior with multiple possible
transitions in the observable part of the GRB remnant's lifetime may
be seen. Detailed, high resolution 3D simulations of a spherical GRB
exploding within the core of a dense stellar environment have been
presented here, which show the resulting dynamics of the remnants
following the onset of the non-relativistic phase.  The resulting
afterglow light curve will depend fairly strongly on the properties of
the system, especially the mass-loss rates of the stars and their
number density with various implications.  On the one hand, it implies
that one cannot be too specific about the times at which we expect to
see transitions in the observed emission. More constructively, if and
when we do see these transitions, they can be fairly constraining on
the properties of the birth sites. If we continue to see the
population of afterglows dominated by high inferred densities when
compared to the IGM, this is support for blast waves in dense stellar
systems, i.e. support for the origin of SGRBs from dynamically formed
compact binaries or collisions.
 
Stars in a dense stellar system interact with each other, both through
their ionizing radiation and through mass, momentum and energy
transfer in their winds. Mass loss leads to recycling of matter into
the cluster wind gas, often with chemical enrichment.  The task of
finding useful progenitor diagnostics is simplified if the pre-burst
evolution leads to a significantly enhanced gas density in the
immediate neighborhood of the burst.  The detection of absorption
signatures associated with the SGRB environment would provide
important clues about the triggering mechanism and the progenitor.

\acknowledgments ER-R was supported in part by the David and Lucile
Packard Foundation, NASA: Swift NNX08AN88G and DOE SciDAC:
DE-FC02-01ER41176. WHL acknowledges financial support from CONACyT
(45845E) and PAPIIT (IN113007) and thanks the Department of Astronomy
and Astrophysics, UC-Santa Cruz for hospitality.

\newpage

\newpage

\newpage

\end{document}